\documentclass[trackchanges,twocolumn]{aastex701}

\newcommand\aastex{AAS\TeX}

\usepackage{comment}
\usepackage{amsmath}
\usepackage{romannum}

\begin{document}

\title{Template \aastex v7.0.1 Article with Examples\footnote{Footnotes can be added to titles}}

\title{Galactic Microquasar and Supernova Remnants Imprinting on Diffuse Neutrino and Gamma-Ray Sky}

\author[0000-0003-0035-7766]{Shiqi Yu}
\email[show]{shiqi.yu@utah.edu}
\affiliation{Department of Physics and Astronomy, University of Utah, Salt Lake City, Utah, USA}

\author[0000-0003-2478-333X]{Bing Theodore Zhang}
\email[show]{zhangbing@ihep.ac.cn}
\affiliation{Key Laboratory of Particle Astrophysics and Experimental Physics Division and Computing Center, Institute of High Energy Physics, Chinese Academy of Sciences, 100049 Beijing, China}
\affiliation{TIANFU Cosmic Ray Research Center, Chengdu, Sichuan, China}

\date{\today}

\begin{abstract}
Recent detections of Galactic diffuse neutrinos by IceCube and $\gamma$-rays by LHAASO offer direct probes into the origin of Galactic cosmic rays. Conventional diffuse templates typically assume a single cosmic-ray injection spectrum across a wide energy range, without accounting for independent contributions from distinct accelerator populations. 
Here, we present a numerical framework that models Galactic diffuse neutrino and $\gamma$-ray emission that incorporates contributions from both microquasar and supernova remnant populations. By anchoring CR injection to local observations and utilizing high-resolution 3D target gas distributions, our model suggests multi-population contributions to the diffuse sky: escaped cosmic rays from supernova remnants dominate below $\sim10\text{ TeV}$, while those from microquasars become the primary driver at higher energies. 
Our predicted neutrino flux agrees well with the recent 12-year IceCube measurements, establishing a physically motivated baseline for the diffuse hadronic background while leaving room for unresolved point-like sources. This flexible framework provides testable predictions for current and future multi-messenger observatories, accommodating diverse accelerator populations and updated observational constraints.
\end{abstract}



\section{Introduction}\label{sec:intro}
The detection of Galactic diffuse neutrino emission at $5.7\sigma$ significance with 12 years of IceCube data \citep{Abbasi:2026fkr} offers a direct window into hadronic cosmic-ray (CR) interactions in the Milky Way. 
To search for this signal, IceCube utilized several diffuse template models constrained by CR and $\gamma$-ray data \citep{Fermi-LAT:2012edv, Gaggero:2015xza, Schwefer:2022zly}. 
However, none of these models account for distinct Galactic source populations with independent injection spectra, nor do they incorporate recent high-precision CR measurements. 
Disentangling point-like sources from the diffuse background remains a central challenge in Galactic multi-messenger astronomy. 
Because Galactic diffuse emission forms the primary background for point-source searches along the Galactic plane, accurately modeling this background is essential for identifying individual neutrino sources and the origin of CRs up to PeV energies. The challenge is underscored by recent IceCube evidence for neutrino sources in the Southern sky near the Galactic disk~\citep{Abbasi:2026oux}.

While supernova remnants (SNRs) have long been considered the primary drivers of Galactic CRs~\citep{Bykov:2025}, they may struggle to accelerate particles to PeV energies~\citep{Bell2013, Cristofari:2020mdf}. 
Microquasars feature relativistic jets and intense winds that provide ideal environments for efficient particle acceleration while they were still active in the past, making them primary candidates as PeV CR accelerators~\citep{LHAASO:2024psv, LHAASO:2025byy,Zhang:2025tew, Wang:2025yqy, Kaci:2025gyb, Aharonian:2026tzf}.

In previous work, we established a detailed Monte Carlo transport framework showing that microquasar remnants could contribute significantly to the CR ``knee'' \citep{Zhang:2026igt}. Building on this pathfinder model, here we extend the framework to multi-messenger observables, presenting a numerical model for Galactic diffuse neutrino and $\gamma$-ray emission that incorporates both supernova remnant and microquasar populations. Concurrently, \cite{Aharonian:2026tzf} demonstrated that minimal $p$/$\text{He}$ components can reproduce local CR observations from GeV to PeV energies. 

Multi-messenger observations of Galactic diffuse neutrinos \citep{IceCube:2023ame, Abbasi:2026fkr} and $\gamma$-rays \citep{LHAASO:2023gne, LHAASO:2024lnz} offer a direct path to test the contributions from these distinct accelerator populations and components. These diffuse secondaries originate from pion decays during hadronic $pp$ interactions with ambient interstellar gas, occurring both near active acceleration sites and across the diffuse Galactic medium.

In this Letter, as a foundational step towards future detailed simulations, we present a novel numerical framework to estimate the diffuse neutrino and $\gamma$-ray emission from Galactic CR accelerator populations. We model Galactic CR transport through a three-dimensional interstellar target gas distribution \citep{Mertsch:2020qld, Mertsch:2022oee}, treating microquasars and supernova remnants as independent source classes to account for their distinct, energy-dependent contributions across the Galaxy. 
By decoupling these source populations rather than assuming a single injection spectrum, our framework incorporates more realistic physical degrees of freedom. Anchored by the latest high-precision local CR measurements, this approach yields an observational baseline testable with high-energy observatories and offers a complementary physical interpretation of the diffuse multi-messenger signatures.

\section{Methodology and Models}\label{sec:method}

\subsection{Source Population and CR Injection}
We simulate the temporal and spatial distribution of source populations following a similar method introduced in~\cite{Zhang:2026igt}. 
Their positions follow a radial density profile \citep[Eq.~4 in][]{Green2015}, while vertical displacements from the Galactic plane are sampled exponentially with a characteristic scale height of $h_z = 0.15\rm~kpc$ ~\citep{Blasi:2011fi}. We evaluate cumulative contributions within a maximum lookback time ($\tau_{\mathrm{lookback}}$, denoting when CR injection started) threshold of $20\rm~Myr$ for both relic microquasars and SNRs. More details regarding population configurations are provided in Appendix~\ref{app:simulation}.
 
For SNRs, we adopt separate injection parameters from~\cite{Aharonian:2026tzf} for proton ($s_{p}=2.40$, $E_{\text{max},\ p}=68$~TeV, $\beta = 2.55$) and helium ($s_{\rm He}=2.24$, $E_{\rm max,\ He}=468$~TeV, $\beta = 1.12$), where $s_{\text{cr}}$ includes diffusion softening correction ($s_{cr}\approx \Gamma - \frac{1}{3}$) and $\beta$ controls the cutoff sharpness. 
This incorporates detailed models that distinguish SNR populations by their helium components, such as metal-rich ejecta or reverse shocks in core-collapse progenitors~\citep{Aharonian:2026tzf}, while demonstrating the flexibility of our framework for future extensions. 
We assume CRs accelerated by SNRs are released instantaneously with a total energy of $\mathcal{E}_{\rm cr}=\eta_{\rm cr, SN}\mathcal{E}_{\rm SN}$, where $\text{cr}\in\{p, He\}$.

For microquasars, we adopt an injection spectral index of $s_{\rm cr} = 2.0$ and a maximum cutoff energy of $E_{\rm max} = 6~\text{PeV}$ for both protons and helium, consistent with parameters inferred from recent studies of GRS 1915+105~\citep{LHAASO:2026gzg} and Cyg X-3~\citep{LHAASO:2025ysm}. 
The total CR luminosity is parameterized via the Eddington luminosity: $L_{\rm cr} = \eta_{\rm cr,MQ} L_{\rm Edd}$, where $L_{\rm Edd} \simeq 1.3 \times 10^{39} (M_{\rm BH}/10 M_\odot) \text{ erg s}^{-1}$  and individual black hole masses are sampled uniformly over $M_{\text{BH}} \in [5.0, 15.0]\ M_\odot$. 
Assuming continuous injection over an active duration $\tau_{\text{dur}}$, the total released CR energy is $W_{\rm cr} = L_{\rm cr}\tau_{\rm dur}$. 

In this work, the efficiencies $\eta_{\rm cr, MQ}$ and $\eta_{\rm cr, SN}$ for both proton and helium components are treated as free parameters and fitted jointly to the observed local CR spectra (> 1TeV) from DAMPE~\citep{DAMPE:2025opn} and LHAASO~\citep{LHAASO:2025byy}, motivated by the primary focus of this study: interpreting high-energy neutrino, $\gamma$-ray, and CR data within a unified multi-messenger framework.

\subsection{CR Density Distribution}
For an individual source with continuous injection at rate $L_{\rm cr}$, the expected CR density contribution $n_{\rm cr}(E, \mathbf{r}, \tau_{\rm lookback})$ at position $\mathbf{r} = (x, y, z)$ is determined by its injection history:
\begin{equation}
n_{\rm cr}(E, \mathbf{r}, \tau_{\rm lookback}) = L_{\rm cr} \int_{\tau_{\min}}^{\tau_{\rm lookback}} \mathcal{P}(\mathbf{r}, E, \tau) \, d\tau \, ,
\label{eq:continuous_injection}
\end{equation}
where $\mathcal{P}(\mathbf{r}, E, \tau)$ represents the spatial diffusion propagator for a particle of energy $E$ after travel time $\tau$, and $\tau_{\min} = \max(0, \tau_{\rm lookback} - \tau_{\rm dur})$. For SNRs, we assume instantaneous escape ($\tau_{\text{dur}} \to 0$), in which case Equation~\ref{eq:continuous_injection} reduces to $n_{\rm cr}=W_{\text{cr}} \mathcal{P}(\mathbf{r}, E, \tau_{\text{lookback}})$, where $W_{\rm CR}$ is the total injected energy. 

To model isotropic diffusion with free escape at the Galactic halo height of $H_G = \pm 4$~kpc, we employ a z-dependent imaging method following~\citet{Blasi:2011fi}. Details regarding the propagator and halo boundary conditions are provided in Appendix~\ref{app:simulation}.

\subsection{Secondary Production}
As these accelerated CRs propagate through the interstellar medium (ISM), they undergo inelastic ($pp$ or He-$p$) collisions with the ambient gas composed of atomic hydrogen H$_{\rm \Romannum{1}}$ \citep{Mertsch:2022oee}, molecular hydrogen H$_2$~\citep{Mertsch:2020qld}, and heavier elements~\citep{Ferriere:2001rg}.

The differential emissivity of secondary particles ($\gamma$-rays and neutrinos) from $pp$ interactions at position $\mathbf{r}$ is given by:
\begin{equation}
q_{\nu}(E_{\nu}, \textbf{r}) = n_{\rm gas}(\textbf{r}) \int dE_{\rm p} n_{\rm p}(E_{\rm p}, \textbf{r}) \beta c \sigma_{pp}\,(E_{\rm p}) \frac{dN_{\nu}}{dE_{\nu}}(E_{\nu}, E_{\rm p}),
\end{equation}
where $n_{\rm gas}(\textbf{r})$ represents the target gas density, $n_{\rm p}(E_{\rm p}, \mathbf{r})$ is the CR proton spatial density, $\beta c$ is the particle velocity, $\sigma_{pp}(E_{\rm p})$ is the inelastic cross-section at proton energy ($E_p$) in the laboratory frame, and $dN_{\nu}/dE_{\nu}$ denotes the differential secondary particle yield from charged pion decays~\citep{Kamae:2006bf}. The ISM helium is scaled by a factor of 0.11 with respect to the total hydrogen nucleon number density~\citep{1998ApJ...509..212S,Evoli:2017vim} and the heavier elements than protons in CR are scaled by 1.1 from the helium components. The contributions of p and $He$ to neutrinos are incorporated inclusively from \textsc{AAfrag}~\citep{Koldobskiy:2021nld}. 

We calculate the secondary neutrino and $\gamma$-ray emission from the proton and helium components of both source populations, whose spectra are jointly fit to local CR measurements. 
For nuclei heavier than helium, we assume a rigidity-dependent CR spectrum. Since the kinetic energy per nucleon scales as $E_k/A \propto (Z/A)\mathcal{R}$, the energy dependence of heavier nuclei is expected to be similar to that of helium nuclei. Therefore, we estimate the neutrino spectrum of heavier nuclei follows the helium-induced spectrum, scaled by their relative abundance to helium ($\simeq$10\%).

Integrating the local emissivity along the line of sight (LOS) yields the differential neutrino flux:
\begin{equation}
\phi_{\nu}(E_{\nu}, \Omega) = \frac{1}{4\pi} \int dE_{\rm p} \, \beta c \, \sigma_{pp}(E_{\rm p}) \, \frac{dN_{\nu}}{dE_{\nu}}(E_{\nu}, E_{\rm p}) \, X(\Omega, E_{\rm p}),
\end{equation}\label{eq:phi_nu}
where $X(\Omega, E_{\rm p}) = \int_0^\infty ds \, n_{\rm gas}(\mathbf{r}) \, n_{\rm p}(E_{\rm p}, \mathbf{r})$ and $\mathbf{r}(s) = \mathbf{r}_\odot + s\mathbf{n}(\Omega)$ accounts for the spatial overlap between CRs and the target gas along the LOS direction $\textbf{n}(\Omega)$ from the solar position $\textbf{r}_\odot$. The full-sky integrated neutrino flux over Galactic longitude ($l$) and latitude ($b$) is then:
\begin{equation}
F_{\nu}(E_{\nu}) = \int_{-\pi/2}^{\pi/2} \int_0^{2\pi} \phi_{\nu}(E_{\nu}, l, b) \cos b \, dl \, db.
\end{equation}

\section{Results}
In our results, we show that neutrinos produced by CRs that had escaped from microquasars contribute significantly to the diffuse neutrino and $\gamma$-ray fluxes above $\sim 10 \text{ TeV}$, while at lower energies, the diffuse emissions are dominated by SNRs.

\subsection{Energy spectrum}

\begin{figure}
    \centering
    \includegraphics[width=\linewidth]{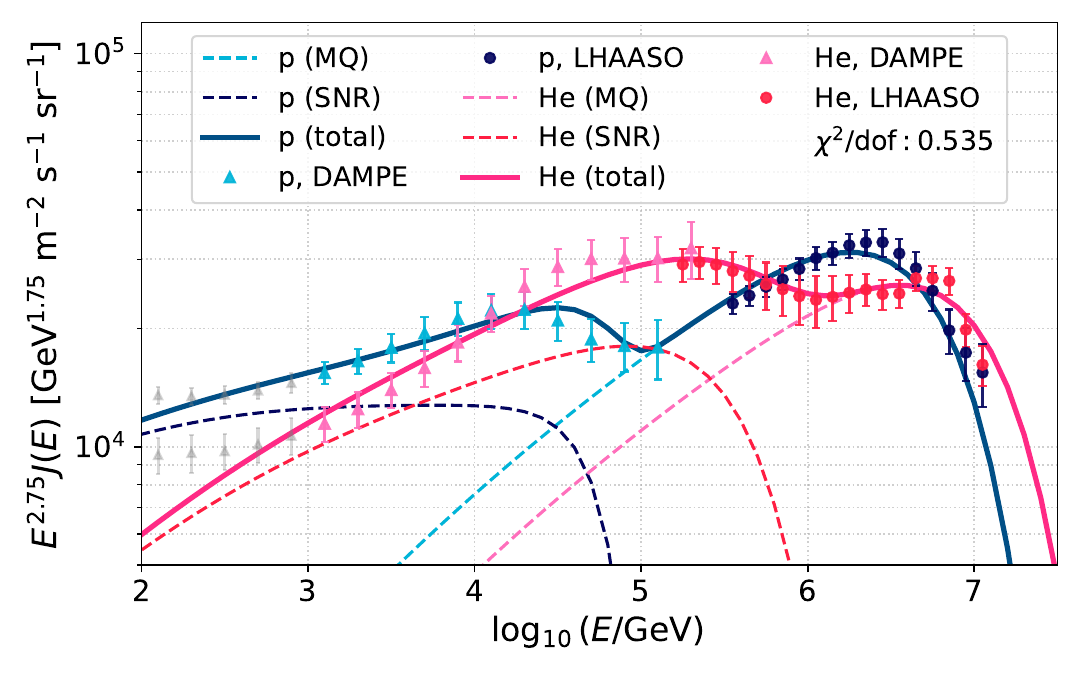}
    \caption{Best-fit spectra for proton (blue) and helium (pink) components from microquasar (light dashed) and SNR (dark dashed) populations, compared against DAMPE (dark spot) \citep{DAMPE:2025opn} and LHAASO (light triangle) \citep{LHAASO:2025byy,LHAASO:2025mlf} observations. Sub-TeV data (gray) are excluded from the fit and displayed for completeness.}
    \label{fig:cr}
\end{figure}

To estimate our secondary neutrino flux, we first constrain the CR acceleration efficiencies $\eta$ by fitting our numerical model to observed CR spectra. The normalization fit incorporates the combined statistical and systematic error bars of the DAMPE and LHAASO observations. The resulting best-fit parameters and their corresponding fit uncertainties are summarized in Table~\ref{tab:cr_fit_results}, with the primary CR spectral comparisons shown in Figure~\ref{fig:cr}.

\begin{table}[htbp]
    \centering
    \caption{Best-fit CR acceleration efficiencies ($\eta$) for the proton and helium components from microquasars and SNRs, with uncertainties reflecting the combined observational errors of the primary CR data.}
    \label{tab:cr_fit_results}
    \begin{tabular}{lcc}
        \hline
        \hline
        Component & Parameter & Fitted Value (\%) \\
        \hline
        Microquasars & $\eta_{p, \text{MQ}}$    & $9.59 \pm 0.18$ \\
                     & $\eta_{\text{He, MQ}}$ & $2.96 \pm 0.08$ \\
        \hline
        SNRs         & $\eta_{p, \text{SN}}$    & $6.00 \pm 0.27$ \\
                     & $\eta_{\text{He, SN}}$ & $2.00 \pm 0.08$ \\
        \hline
        \hline
    \end{tabular}
\end{table}

For microquasar-induced protons, the required acceleration efficiency is $\eta_{p, \text{MQ}} \sim 10\%$, slightly higher than our previous estimate ($\eta_{p, \text{MQ}} \sim 3\%$; \citealt{Zhang:2026igt}), likely due to Galactic magnetic field (GMF) transport and the influence of nearby microquasars in our MC framework. Notably, because local CR observations are sensitive to nearby individual sources~\citep{Zhang:2026igt}, the fitted efficiencies $\eta$ strictly reflect the local normalization. Combining these values with our modeled source distributions yields our baseline secondary flux predictions.

\begin{figure}[htbp]
\centering
\includegraphics[width=\linewidth]{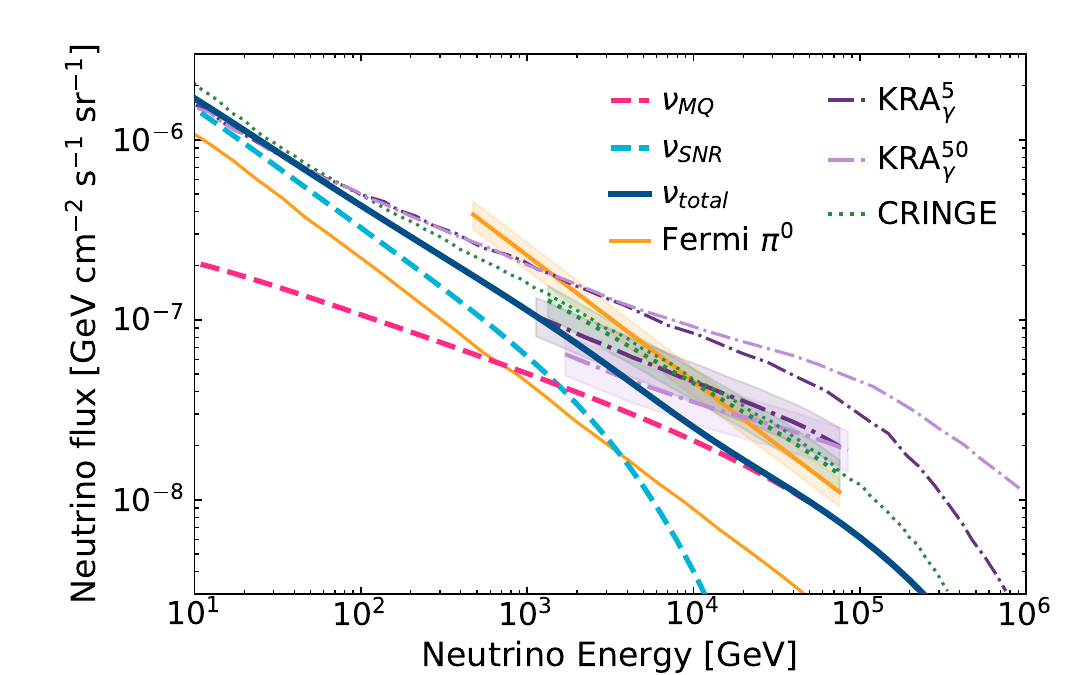}
\caption{Total predicted per-flavor Galactic diffuse neutrino spectra for the inner Galaxy (dark blue), decomposed into microquasar (light pink) and SNR (light blue) populations, compared with IceCube measurements and nominal spectra of Fermi-$\pi^0$, {\sc CRINGE}, KRA$_\gamma^5$ and KRA$_\gamma^{50}$ models~\citep{Abbasi:2026fkr}.
}
\label{fig:neutrino_spectrum}
\end{figure}

\begin{figure}
    \centering
    \includegraphics[width=\linewidth]{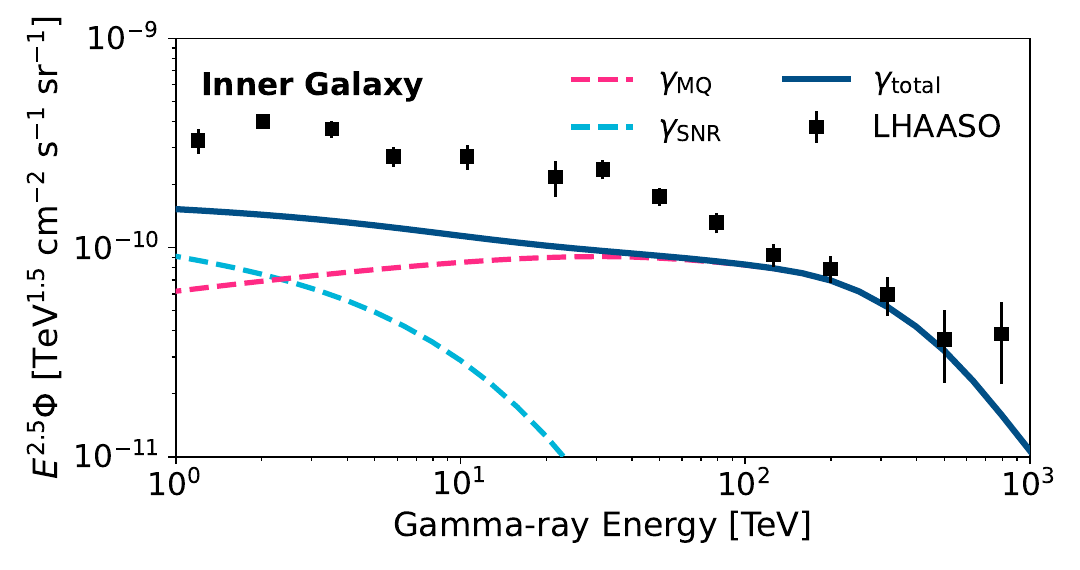}
    \includegraphics[width=\linewidth]{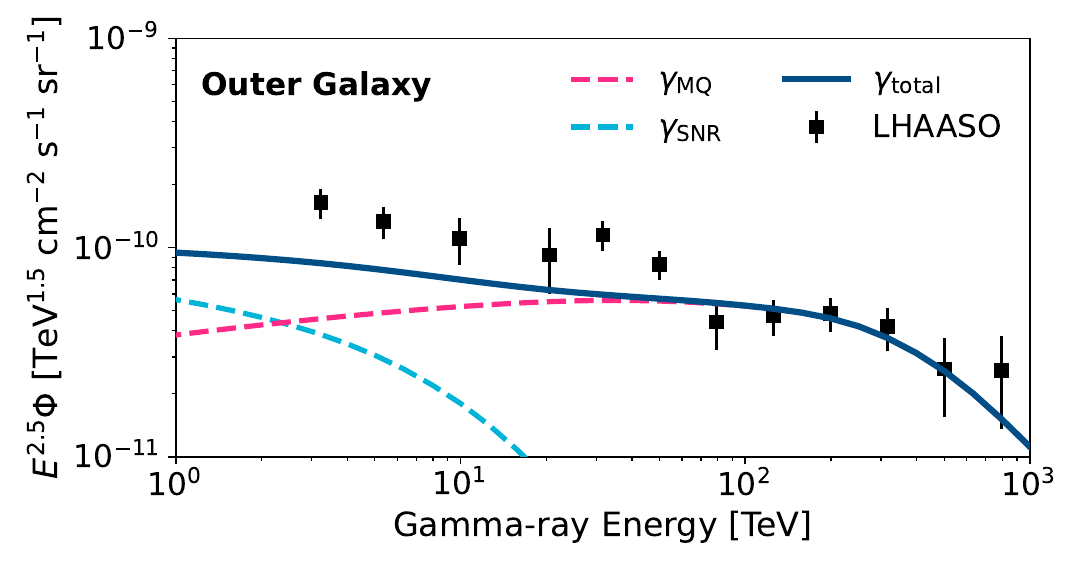}
   \caption{Diffuse gamma-ray flux integrated over the inner (top) and outer (bottom) Galactic regions within $|b|\leq5^\circ$, compared with LHAASO measurements. The same sky mask as in~\cite{LHAASO:2024lnz} is applied.}
    \label{fig:diffuse-gamma-nu-LHAASO}
\end{figure}

Incorporating these fitted $\eta_{cr}$ values, we predict LOS per-flavor diffuse neutrino spectra from both populations integrated over the inner sky ($|b|<15^\circ, |l|<20^\circ$), shown in Fig.~\ref{fig:neutrino_spectrum}. For secondary flux predictions (Figures~\ref{fig:neutrino_spectrum} and \ref{fig:diffuse-gamma-nu-LHAASO}), we present nominal central curves, as local normalization uncertainties on $\eta$ are negligible within our numerical framework~\citep{Zhang:2026igt}. Diffuse neutrino measurements are currently provided as template fits rather than model-independent unfolded flux points; we therefore compare our model against these nominal template predictions and their IceCube best fits~\citep{Abbasi:2026fkr}. A quantitative comparison to data points will be essential in future studies.

The inner Galaxy serves as a stringent testing ground due to its high sensitivity to target gas distributions and source spatial distributions. Without any fine-tuning, our predicted spectrum, anchored directly on local CR observations, aligns closely with existing predictions, particularly in the sub-TeV regime where SNRs dominate in our model. This sub-TeV agreement arises because low-energy CR observations broadly anchor the SNR contribution across models. Even though other templates adopt different injection spectra without assuming independent source distributions, they are all tuned to GeV–TeV CR measurements and are therefore constrained to the same order of magnitude.

Above $\sim 10\text{ TeV}$, the microquasar-dominated spectral hardening aligns well with the general trend of the \textsc{CRINGE} and $\text{KRA}_\gamma$ templates, driven by the hard injection spectra of the microquasar population as suggested by our model. However, because the $\text{KRA}_\gamma$ model \citep{Gaggero:2015xza} was not constrained against the latest TeV to PeV CR measurements, its high-energy extrapolation yields a noticeably higher flux. Our predictions agree more closely with \textsc{CRINGE} \citep{Schwefer:2022zly}, which incorporates multiple rigidity-dependent diffusion breaks. However, their reliance on a single injection spectrum differs fundamentally from our multi-population framework, where source acceleration and maximum cutoffs physically dictate the spectral shape and its eventual steepening.

Neither source population alone reproduces the Galactic diffuse neutrino data across all energies. In contrast, their combined emission matches existing template predictions, yielding a smooth transition across different energy regimes and a natural prediction of the spectral hardening in neutrino data above 10~TeV. Notably, our central model predictions remain below observation bounds, leaving room for potential additions such as faint unresolved sources or nearby active accelerators.

The predicted diffuse $\gamma$-ray flux incorporates energy-dependent attenuation via pair production ($\gamma\gamma \to e^+e^-$) on the cosmic microwave background (CMB), extragalactic background light (EBL)~\citep{Gilmore:2011ks}, and the interstellar radiation field~\citep{Robitaille:2012kg}. As shown in Fig.~\ref{fig:diffuse-gamma-nu-LHAASO}, our predicted diffuse $\gamma$-ray fluxes in both the inner and outer Galactic regions agree well with LHAASO observations~\citep{LHAASO:2023gne, LHAASO:2024lnz}. 

Below $\sim100~\text{TeV}$, a residual excess remains in the $\gamma$-ray data relative to our hadronic prediction, likely originating from unmodeled Inverse Compton scattering by CR electrons~\citep{Yan:2023uxd, Kaci:2024lwx, Das:2026vlj} as well as unresolved, faint sources~\citep{LHAASO:2024lnz, Fang:2023ffx,Marinos:2025ddy,Menchiari:2026ney}. Above $\sim100\text{ TeV}$, however, Inverse Compton is strongly suppressed by Klein-Nishina effects, transitioning the diffuse $\gamma$-ray sky into a purely hadronic regime dominated by our hard microquasar proton spectrum, which naturally reproduces the high-energy LHAASO data. 
Combining high-resolution diffuse $\gamma$-ray maps with unattenuated neutrino data provides a promising path forward to disentangle sub-hundred-TeV leptonic residuals and characterize the PeV Galactic hadronic emission, representing a key application for our flexible multi-population framework.

\subsection{2D Spatial Diffuse Maps}

\begin{figure*}[bth]
    \centering
    \includegraphics[width=0.49\textwidth]{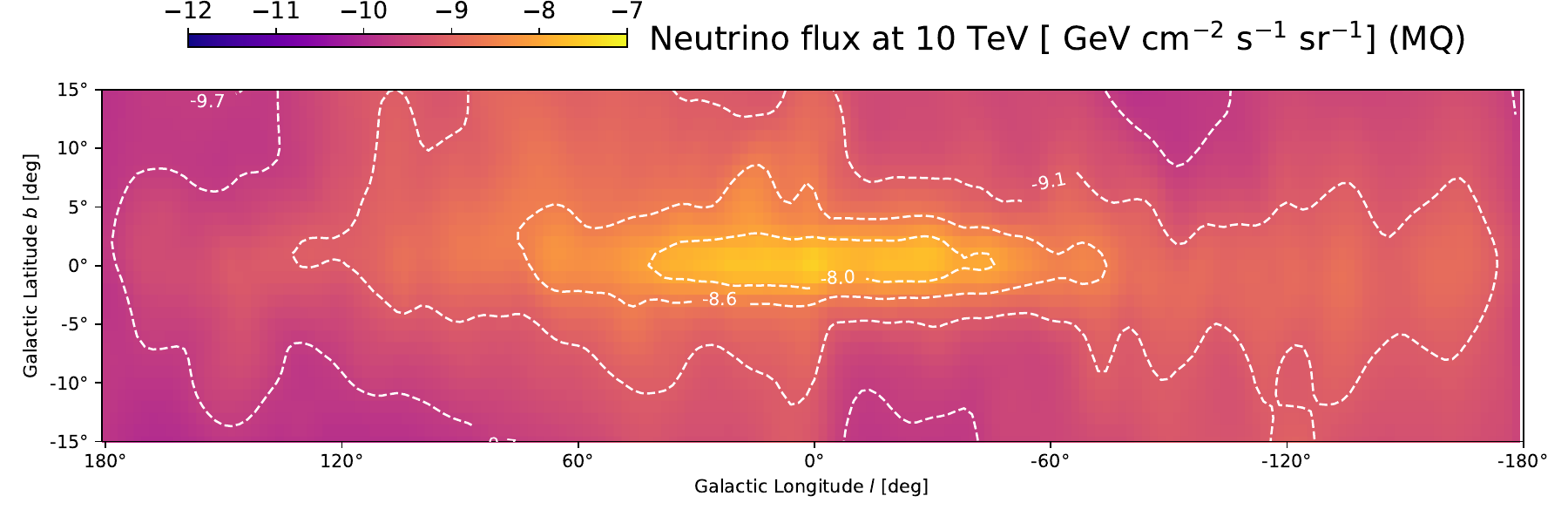}
    \includegraphics[width=0.49\textwidth]{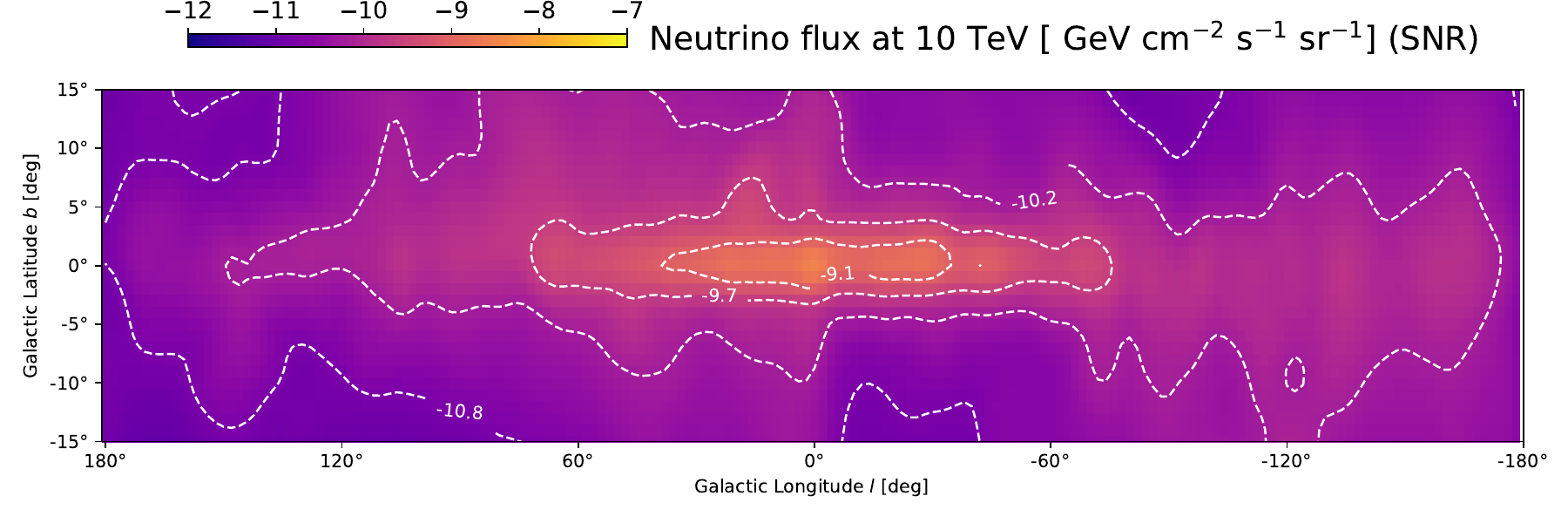}
    \includegraphics[width=0.49\textwidth]{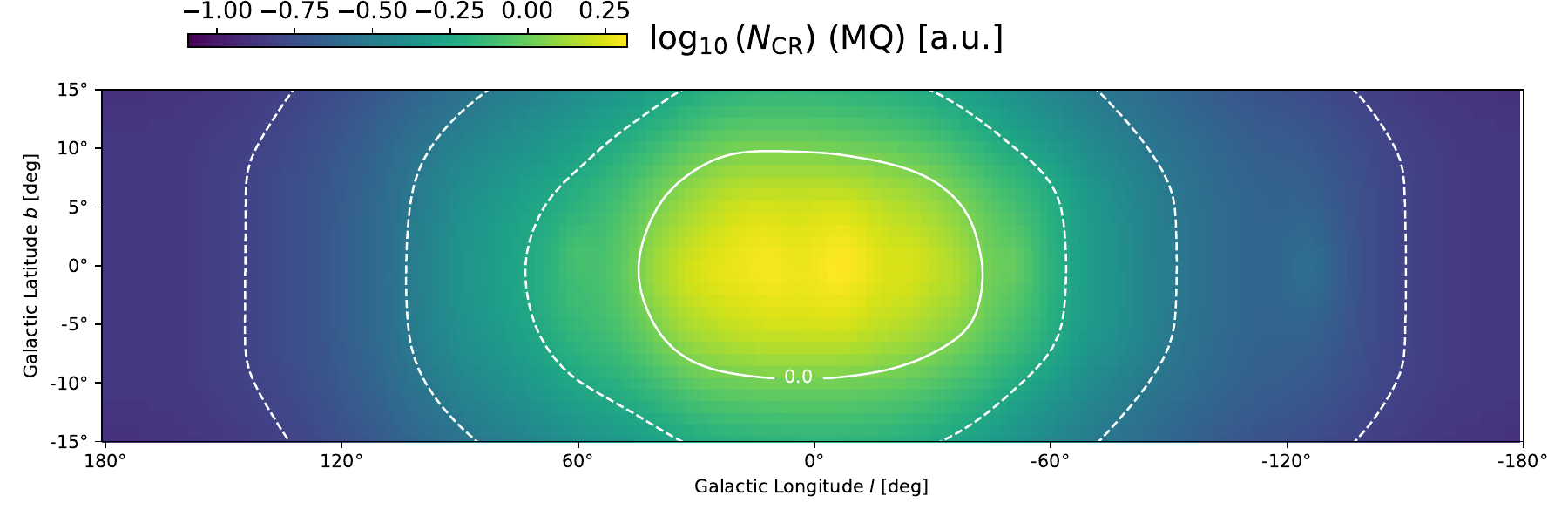}
    \includegraphics[width=0.49\textwidth]{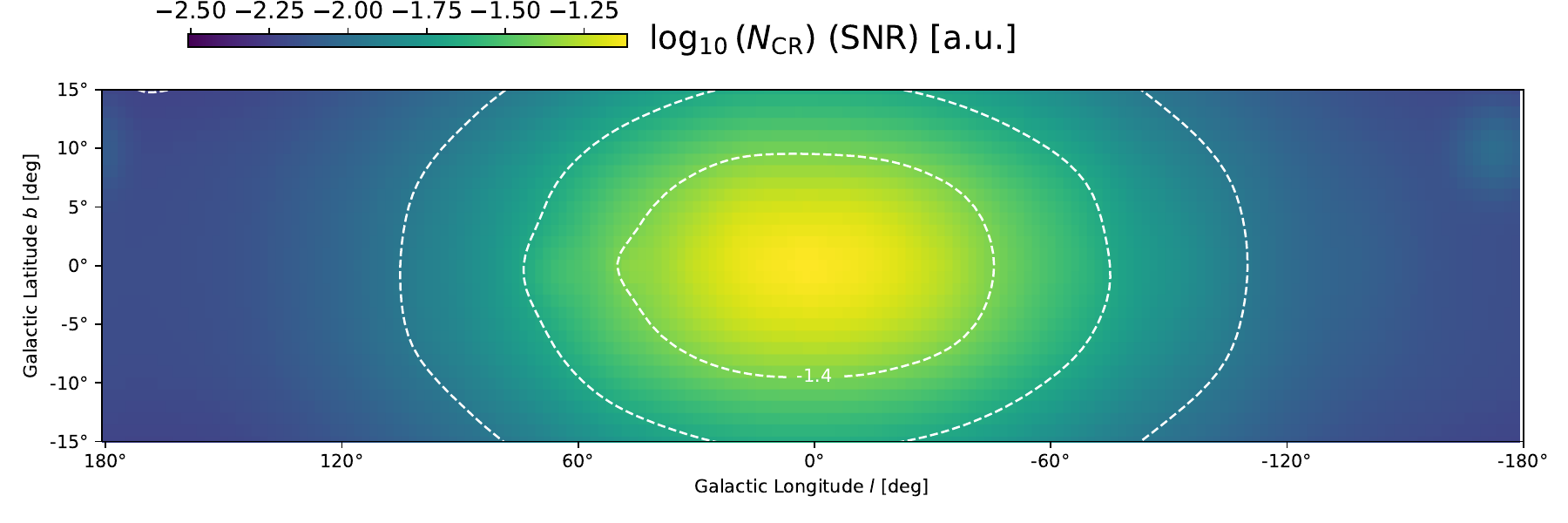}    \includegraphics[width=0.49\textwidth]{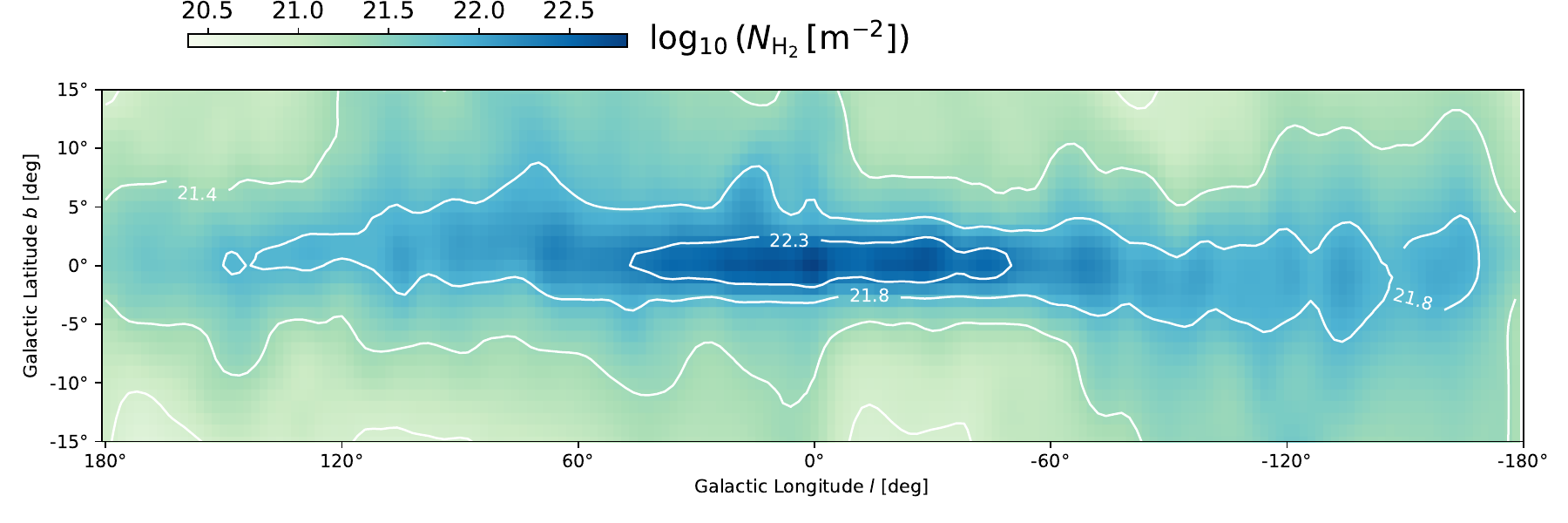}
    \includegraphics[width=0.49\textwidth]{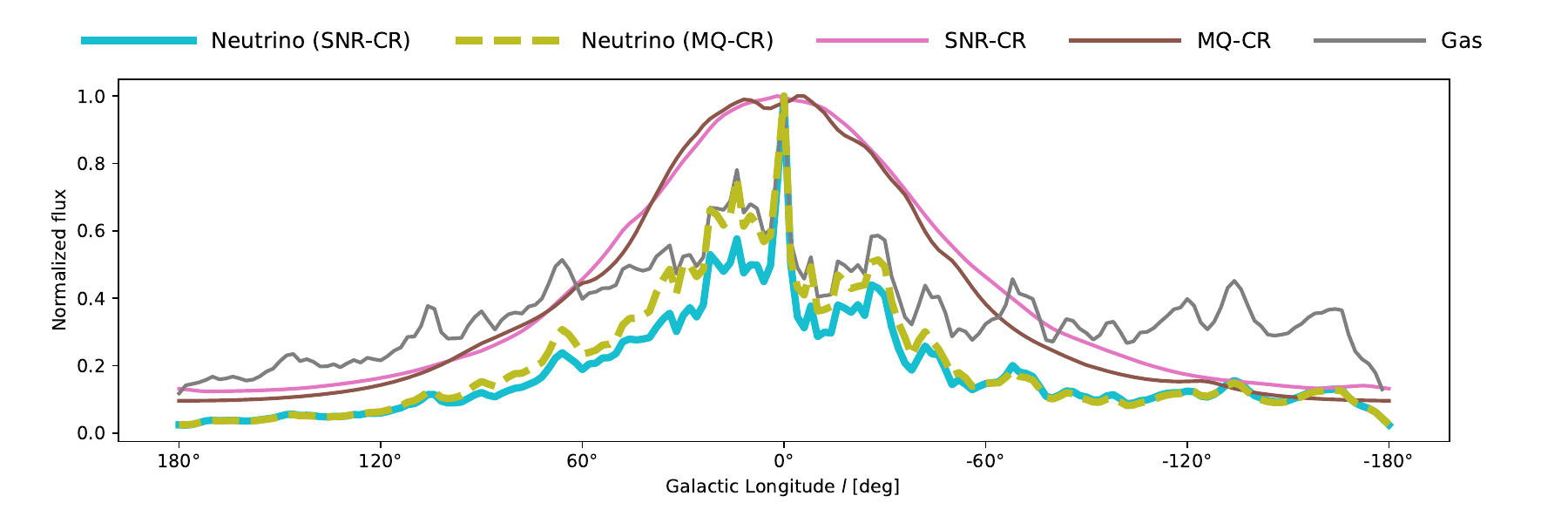}
    \caption{Projected distributions in Galactic coordinates ($|l|<180^\circ$, $|b|<15^\circ$) predicted for microquasar (left column) and SNR (right column) populations associated with induced CR protons. The top row shows the neutrino flux emissivity templates at 10 TeV. The middle row shows the CR density distributions with arbitrary normalization. The bottom-left panel shows the interstellar gas density distribution. The bottom-right panel summarizes the corresponding 1D longitude profiles integrated over $|b|<15^\circ$. All maps are smoothed with a Gaussian kernel of $\sigma=1.1^\circ$. }
    \label{fig:GalacticPlane}
\end{figure*}

Under our assumption of isotropic spatial diffusion, historically injected CRs slowly accumulate to form a broad, continuous background across the Galactic plane that smoothly mirrors the large-scale radial distribution of each source population (middle row of Fig.~\ref{fig:GalacticPlane}). 
However, because neutrino production requires hadronic interactions with target matter, the resulting secondary flux maps inherit the complex, filamentary structure of the interstellar target gas (bottom-left panel of Fig.~\ref{fig:GalacticPlane}). 

Consequently, localized ``hot spots" in neutrino skymaps need not correspond to individual active accelerators; rather, they reveal LOS where the steady CR sea imprints onto dense target gas. The 1D longitudinal profiles integrated over $|b|<15^\circ$ (bottom-right panel of Fig.~\ref{fig:GalacticPlane}) illustrate this clearly: the secondary neutrino profiles trace the sharp spikes of the target gas, showing that local mass concentrations govern the fine-spatial structures. On broader scales, the underlying CR modulates this distribution, producing a more pronounced inner-Galaxy peak in $l$, than observed in the gas profile itself.

\section{Discussion}
Our model evaluates the Galactic diffuse multi-messenger emission by incorporating two independent populations that dominate distinct energy regimes. Our predicted neutrino spectrum broadly agrees with empirical templates, with the overall spectral shape governed by the distinct injection indices and maximum acceleration energy cutoffs of each source class. 
In particular, our framework demonstrates that a spectral hardening above $10\text{ TeV}$ in the diffuse emission naturally emerges as the microquasar population takes over above the softer SNR component.

Our predictions also align well with the observed high-energy diffuse $\gamma$-ray sky. Below $\sim100\text{ TeV}$, unmodeled leptonic Inverse Compton scattering and faint unresolved source populations complicate the diffuse $\gamma$-ray flux, creating residual excess relative to pure hadronic models. Above $\sim100\text{ TeV}$, Klein-Nishina suppression of leptonic emission effectively makes this background negligible, leaving a clean hadronic baseline. In this regime, our model enables a direct association between PeV $\gamma$-rays and high-energy diffuse neutrinos dominated by microquasar PeVatrons. 

Connecting this global hadronic emission to local observables, however, introduces a key mapping challenge. The local CR flux is inherently subject to stochastic fluctuations from nearby source distributions and the local GMF, especially above 100~TeV due to the lower spatial density of microquasars~\citep{Zhang:2026igt}. 
Consequently, while local CR measurements provide a valuable constraint on the overall flux normalization, they do not reflect regional variations across the broader Galaxy. 
A key element of our framework is using the modeled spatial source distribution to globally extrapolate locally constrained CR efficiencies ($\eta$). Although random source realizations introduce spatial flux variations, this systematic uncertainty can be quantified using Monte Carlo ensembles and tested against independent observations with neutrinos and $\gamma$ rays. 

As proposed in \citet{Zhang:2026igt}, CRs accelerated by recent or active sources may remain temporarily confined by local magnetic environments, having not yet reached the Solar neighborhood. Consequently, while these active accelerators do not contribute to the observed local CR data, their primary particles continuously interact with surrounding target gas, leaving additional imprints beyond the steady-state CR sea directly on diffuse neutrino skymaps.

Future work can systematically evaluate the additional impacts of hadronic interaction models, detailed interstellar gas maps, and GMF transport parameters. Nevertheless, our present framework establishes a robust, physically grounded baseline for evaluating diffuse backgrounds, essential for isolating both resolved and unresolved point sources.

As multi-messenger observatories continue to resolve the Galactic plane at sub-PeV energies, incorporating population-level source diversity will be essential for disentangling source physics from interstellar CR transport, providing a testable template for current and future high-statistics measurements with IceCube, KM3NeT, and Baikal-GVD. Studies of point-like sources in the region of the Galactic plane further require a critical understanding of Galactic diffuse emission. While our present framework establishes a robust baseline, achieving fully self-consistent templates of high-energy Galactic diffuse emission will require incorporating anisotropic transport, source distribution fluctuations, and photon absorption, which represent the immediate next steps of our work.

\section{Summary}
In this study, we have presented a multi-population framework to evaluate the Galactic diffuse neutrino and accompanying $\gamma$-ray fluxes. 
By calculating the diffuse CR density released from microquasars and SNRs from the past and mapping it onto a realistic 3D interstellar gas distribution, we demonstrated that the Galactic diffuse spectrum could be naturally decomposed by them. 

While SNRs dominate the lower-energy spectrum in agreement with standard diffuse neutrino templates, the hard CR injection from the microquasar population introduces a distinct spectral hardening extending to PeV energies. 
The microquasar component establishes a hadronic baseline that reproduces LHAASO observations above $\sim100\text{ TeV}$, while lower-energy $\gamma$-ray residuals suggest contributions from unresolved sources and leptonic processes.

By extrapolating locally constrained CR acceleration efficiencies assuming source population distributions for global LOS fluxes, our framework can account for the local stochasticity of low-density source populations and provide testable multi-messenger templates. Because Galactic diffuse emission forms the dominant background for point-source searches along the Galactic plane, establishing this physically grounded baseline provides testable templates for current and next-generation multi-messenger observatories to disentangle PeVatrons from diffuse backgrounds. 

\appendix

\section{Simulation Setup}\label{app:simulation}

\begin{table}[htb]
\caption{Parameters for Population Distributions and CR Injection of microquasars and SNRs.}
\label{tab:pop_parameters}
\centering
\begin{tabular}{lcc}
\hline
\hline
Source & Microquasar & SNR\\
\hline
\multicolumn{3}{c}{\textbf{Shared Parameters}} \\
\hline
Radial Distribution Parameters & \multicolumn{2}{c}{$\alpha = 1.09, \quad \beta = 3.87$} \\
Pivot Height ($h_z$)  & \multicolumn{2}{c}{$0.15\rm~kpc$} \\
Max.\ Lookback & \multicolumn{2}{c}{$20\rm~Myr$} \\
\hline
\multicolumn{3}{c}{\textbf{Distinct Parameters}} \\
\hline
Source Rate & $1 \times 10^{-4} \text{ yr}^{-1}$ & $1/30 \text{ yr}^{-1}$\footnote{\cite{Blasi:2011fi}.}\\
Jet Duration ($\tau_{\rm dur}$) & Uniform($0.1, 1.0\rm~Myr$) & -- \\
CR Injection Index ($s_{\rm cr}$) & 2.0 & 2.4 (2.24) for $p$ (He) \\
CR Max.\ Accelerating Energy ($E_{\rm max}$) & $6\rm~PeV$ & $68 (468)\rm~TeV$ for $p$ (He)\\
Mass / Kinetic Energetics & $M_{\rm BH} \sim \text{Uniform}(5.0, 15.0) \, M_\odot$ & $\frac{E_{\rm SN}}{\rm erg} \sim$ Log-Norm(50.5, 0.53$^2$)\footnote{\cite{2022Univ....8..653L}.} \\
\hline
\hline
\end{tabular}
\end{table}

In this appendix, we detail the configurations of our numerical framework, as summarized in Table.~\ref{tab:pop_parameters}. The radial density distribution of SNRs as a function of Galactocentric radius $r$ is parameterized as: \begin{equation}f(r) = \left(\frac{r}{r_\odot}\right)^\alpha \exp\left(-\beta \frac{r-r_\odot}{r_\odot}\right),\end{equation}
where $r_\odot = 8.5\text{ kpc}$ is the solar Galactocentric distance~\citep{Green2015}. To the vertical height $z$ of the source is sampled from a exponential distribution:\begin{equation}P(z) = \frac{1}{2h_z} \exp\left(-\frac{\vert z\vert}{h_z}\right),\end{equation}
where $h_z = 0.15\text{ kpc}$ denotes the characteristic scale height of the thin disk.

For CRs $> 1\text{ TeV}$, diffusive escape ($\tau_{\text{esc}} \lesssim 1\text{ Myr}$) dominates over hadronic losses ($\tau_{\text{loss}} \sim 50 \, [n_{\text{gas}} / 1\text{ cm}^{-3}]^{-1}\text{ Myr}$). Therefore, a maximum lookback time threshold of $20\rm~Myr$ is applied for computational efficiency, which comfortably exceeds the required time window for reaching a steady-state CR ``sea" above TeV. Extending the lookback window further yields negligible change. The lookback time of each source is sampled uniformly within this range.

The acceleration phase of an SNR is relatively short compared to multi-million-year Galactic transport timescales; we therefore treat SNR injections as instantaneous burst events occurring at $\tau_{\mathrm{lookback}}$.

Adopting the parameterization from~\cite{Leahy:2020jpk}, the total kinetic energy released in an individual SN burst, $E_{\text{SN}}$, is sampled from a log-normal distribution:
\begin{equation}
\log_{10}\left(\frac{E_{\text{SN}}}{1\text{ erg}}\right) \sim \mathcal{N}(50.5, 0.53^2).
\end{equation}

For the microquasar population, active jet durations $\tau_{\rm dur}$ are sampled uniformly between $0.1\text{ Myr}$ and $1.0\text{ Myr}$, during which microquasars inject CRs at a constant rate $L_{\rm cr}$. Protons and helium are injected separately, with helium serving as a proxy for the total heavier nuclei flux to properly account for rigidity-dependent transport.

Inspired by the image method in~\citep{Blasi:2011fi}, we setup our spatial propagator as:
\begin{equation}
\mathcal{P}(\mathbf{r}, z, E, \tau) = \frac{1}{(4\pi \, D(E) \, \tau)^{3/2}} \, \exp\left( -\frac{r^2}{4 D(E) \tau} \right) \,\mathcal{V}(z, \tau, E) \, ,
\label{eq:blasi_kernel}
\end{equation}
where $r = \sqrt{(x-x_s)^2 + (y-y_s)^2}$ is 2D radial distance to the source in the Galactic plane, and $D(E) = D_0 (E / E_0)^\delta$ is the rigidity-dependent spatial diffusion coefficient with $D_0 = 6 \times 10^{28}\text{ cm}^2\text{ s}^{-1}$, $E_0 = 4\text{ GV}$, and index $\delta = 1/3$ assuming slow diffusion~\citep{Blasi:2011fi, Zhang:2026igt}. The free-escape boundary conditions at $H = \pm 4$~kpc are enforced by:
\begin{equation}
\mathcal{V}(z, \tau, E) = \sum_{n=-\infty}^{+\infty} (-1)^n \exp\left( -\frac{(z-z'_n)^2}{4 D(E) \tau} \right),
\end{equation}
where $z'_n = z_s + 2nH$ represents the vertical coordinates of the image sources needed to ensure $\mathcal{P}(z=\pm H, E, \tau) = 0$ for a real source located at $z_s$.


\bibliography{sample701}{}
\bibliographystyle{aasjournalv7}



\end{document}